\documentclass[10pt, preprint2]{aastex}

\newcommand{\vz}{\mbox{$v_{\rm Z}$}}
\newcommand{\vrot}{\mbox{$v_{\rm rot}$}}
\newcommand{\vphi}{\mbox{$v_\varphi$}}
\newcommand{\wz}{\mbox{$w_{\rm z}$}}
\newcommand{\wzz}{\mbox{$w_{\rm z}^2$}}

\newcommand{\taul}{\mbox{$\tau_0$}}

\newcommand{\BHan}{\mbox{$B_{\rm Han}$}}
\newcommand{\alt}{\mbox{$\alpha_2$}}
\newcommand{\psis}{\mbox{$\psi_{\rm s}$}}
\newcommand{\thetas}{\mbox{$\theta_{\rm s}$}}
\newcommand{\mustar}{\mbox{$\mu_\ast$}}

\newcommand{\fiso}{\mbox{${\cal F}_\nu^{\rm iso}$}}
\newcommand{\qs}{\mbox{$q_\nu^{\rm s}$}}
\newcommand{\us}{\mbox{$u_\nu^{\rm s}$}}

\newcommand{\phat}{\mbox{$\hat{\varphi}$}}
\newcommand{\Zhat}{\mbox{$\hat{Z}$}}

\newcommand{\hanvec}{\mbox{$\vec{h}$}}
\newcommand{\hI}{\mbox{$h_I$}}
\newcommand{\hQ}{\mbox{$h_Q$}}
\newcommand{\hU}{\mbox{$h_U$}}
\newcommand{\hV}{\mbox{$h_V$}}
\newcommand{\FI}{\mbox{${\cal F}_I$}}
\newcommand{\FQ}{\mbox{${\cal F}_Q$}}
\newcommand{\FU}{\mbox{${\cal F}_U$}}

\def \etal{{\it et~al.\/}}

\begin{document}

\title{The Hanle Effect as a Diagnostic of Magnetic Fields in Stellar
Envelopes. V. Thin Lines from Keplerian Disks }

\author{R.\ Ignace}
\affil{
        Department of Physics \& Astronomy,
        East Tennessee State University,
        Johnson City, TN 37614, USA}

\email{ignace@etsu.edu}

\begin{abstract}

This paper focuses on the polarized profiles of resonance scattering
lines that form in magnetized disks.  Optically thin lines from
Keplerian planar disks are considered.  Model line profiles are
calculated for simple field topologies of axial fields (i.e.,
vertical to the disk plane) and toroidal fields (i.e., purely
azimuthal).  A scheme for discerning field strengths and geometries
in disks is developed based on Stokes Q-U diagrams for the run of
polarization across line profiles that are Doppler broadened by the
disk rotation.  A discussion of the Hanle effect for magnetized
disks in which the magnetorotational instability (MRI) is operating
is also presented.  Given that the MRI has a tendency to mix the
vector field orientation, it may be difficult to detect the disk
fields with the longitudinal Zeeman effect, since the amplitude of
the circularly polarized signal scales with the net magnetic flux
in the direction of the observer.  The Hanle effect does not suffer
from this impediment, and so a multi-line analysis could
be used to constrain field strengths in disks dominated by the MRI.

\end{abstract}


\section{INTRODUCTION}	\label{sec:intro}

Spectropolarimetry continues to be a valuable technique for a broad
range of astrophysical studies (e.g., Adamson \etal\ 2005;
Bastian 2010), including applications for
circumstellar envelopes.  Advances in technology and access to
larger telescopes means an ever growing body of high quality
spectropolarimetric data.  It is therefore important that the arsenal
of diagnostic methods and theoretical models in different astrophysical
scenarios keep pace.   This paper represents the fifth in a series
devoted toward developing the Hanle effect as tool for measuring
magnetic fields in circumstellar media from resonance line scattering
polarization.  The observational requirements for the effects
examined in this series are ambitious:  high signal-to-noise (S/N)
data and high spectral resolving power.  However, these demands are
being met, as illustrated by Harrington \& Kuhn (2009a) in a
spectropolarimetric survey of circumstellar disks at H$\alpha$.

There are numerous effects that can influence the polarization
across resolved lines.  A number of researchers have investigated
the effects of line opacity for polarization from Thomson scattering
(Wood, Brown, \& Fox 1993; Harries 2000; Vink, Harries, Drew 2005;
Wang \& Wheeler 2008; Hole, Kasen, \& Nordsieck 2010).  Scattering
by resonance lines can generate polarization similar to dipole
scattering (e.g., Ignace 2000a).  With high S/N and high spectral
resolution, Harrington \& Kuhn (2009b) have identified a new
polarizing effect for lines that coincides with line absorption.
An explanation for this previously unobserved effect in stars is
discussed by Kuhn \etal\ (2007) and is attributed to the same
dichroic processes detailed by Trujillo Bueno \& Landi Degl'Innocenti
(1997) for interpreting polarizations in some solar spectral lines.
Generally associated with circular polarization, the Zeeman effect
has received acute attention of late as a result of techniques that
co-add many lines (Donati \etal\ 1997).  The method has been used
successfully in many studies (see the review of Donati \& Landstreet
2009).  In relation to massive stars, the technique has led to the
detection of magnetism in several stars (e.g., Donati \etal\ 2002,
2006a, 2006b; Neiner \etal\ 2003; Grunhut \etal\ 2009).  In fact,
there exists a large collaborative effort called the
MiMeS\footnote{www.physics.queensu.ca/~wade/mimes} collaboration
(e.g., Wade \etal\ 2009) to increase the sample of massive stars
with well-studied magnetic fields.

Another method that has been used productively in studies of solar
magnetic fields involves the Hanle effect (Hanle 1924).  This is a
weak Zeeman effect that operates when the Zeeman splitting is roughly
comparable to the natural line broadening.  Monographs that deal
with atomic physics and polarized radiative transfer provide excellent
treatments of the Hanle effect, including Stenflo (1994) and Landi
Degl'Innocenti \& Landolfi (2004).  This contribution extends a
series of theoretical papers that has enlarged the scope of its
general use with circumstellar envelopes.  Building on work developed
in the solar physics community, and using expressions for resonance
line scattering with the Hanle effect (e.g., from Stenflo 1994),
Ignace, Nordsieck, \& Cassinelli (1997; Paper~I) provided an
introduction of the Hanle effect for use in studies of circumstellar
envelopes.  Ignace, Cassinelli, \& Nordsieck (1999, Paper~II)
considered its use in simplified models of disks with constant
radial expansion or solid body rotation as pedagogic examples.  Both
of these papers approximated the star as a point source of illumination.
Ignace (2001; Paper~III) then incorporated the finite source
depolarization factor of Cassinelli, Nordsieck, \& Murison (1987)
into the treatment of the source functions for the Hanle effect in
circumstellar envelopes.  Finally, Ignace, Nordsieck, \& Cassinelli
(2004, Paper~IV) calculated the Hanle effect in P~Cygni wind lines
using an approximate treatment for line optical depth effects.

The focus of this fifth paper in the series is the Hanle effect in
polarized lines from Keplerian disks, of relevance for accretion
disks and Be~star disks (e.g., Cranmer 2009).  As in previous papers
of this series, the Sobolev approximation (Sobolev 1960) for optically
thin scattering lines is adopted for exploring the run of polarization
with velocity shift across line profiles.  This paper adopts the
notation laid out in Paper~II.  The results of Paper~II are expanded
for Keplerian rotation with the inclusion of finite source
depolarization (ala Paper~III) and stellar occultation.  Resonance
line scattering polarization from such disks were explored in Ignace
(2000a) for unmagnetized disks without consideration
of the Hanle effect.

\begin{figure}[t]
\plotone{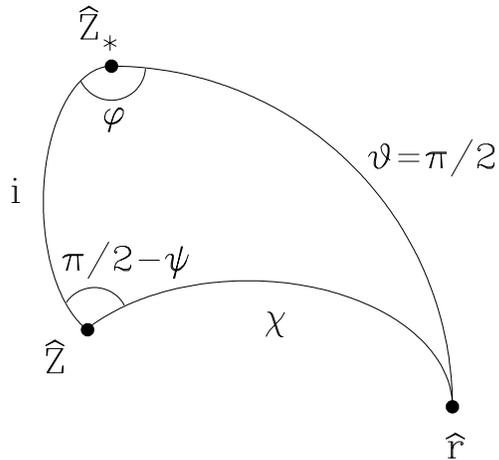}
\caption{Illustration of the relation between observer and stellar
coordinate reference frames.  The observer is in the direction
of $\hat{Z}$; the symmetry axis for the disk is $\hat{Z}_\ast$.
A scattering point is in direction $\hat{r}$.  This point lies
in the equatorial disk, hence $\varphi=\pi/2$.
\label{fig1}}
\end{figure}

Drawing on these previous works, methods for computing the polarization
of line profiles will be briefly reviewed in Section~\ref{sec:lines}.
In Section~\ref{sec:hanle}, results for the Hanle effect from disk
lines are described.  There are three applications that will be
considered: purely axial fields (i.e., normal to the disk plane),
purely toroidal fields (i.e., azimuthal in the disk plane), and
field topologies that can arise from the magnetorotational instability,
or MRI (Balbus \& Hawley 1991).  Conclusions of this study and
observational prospects are presented in Section~\ref{sec:conc}.
Appendices detail analytic derivations for special cases of the
Hanle effect in Keplerian disks.

\section{THIN SCATTERING LINES IN KEPLERIAN DISKS}	\label{sec:lines}

\subsection{Coordinate Systems of the Model}

The focus of this paper is thin scattering lines from a planar
circumstellar disk in which the gas obeys circular Keplerian motion.
To describe this geometry and the line scattering polarization, a
number of coordinates will be needed, which are introduced
here.

\begin{itemize}

\item A Cartesian coordinate system for the star is assigned to be
$(X_\ast, Y_\ast, Z_\ast)$.  The $Z_\ast$ axis is the rotation axis
of the star and disk.

\item Another one for the observer is $(X,Y,Z)$.  The observer is
located at large positive distance along the $Z$ axis.

\item Spherical coordinates in the star frame are $(r,\vartheta,
\varphi)$.  Cylindrical coordinates are identified by
$(\varpi, \varphi, Z_\ast)$.

\item Cylindrical coordinates in the observer frame are $(p, \psi,
Z)$.

\item Spherical angular coordinates defined in a frame of the local
magnetic field at any point will be $(\theta,\phi)$.  The latter
system is needed to evaluate the Hanle effect.

\item The $Z$ axis is taken to be inclined to the $Z_\ast$ axis by
an angle $i$.

\item The circumstellar disk assumed to be axisymmetric and exists
entirely in the equatorial plane of the star, and so it is located
at $\vartheta=\pi/2$.  The scattering angle between a point in the 
disk to the observer is signified by $\chi$.

\end{itemize}

Figure~\ref{fig1} shows a spherical triangle used in relating the
observer and star coordinate systems.  

\begin{figure}[t]
\plotone{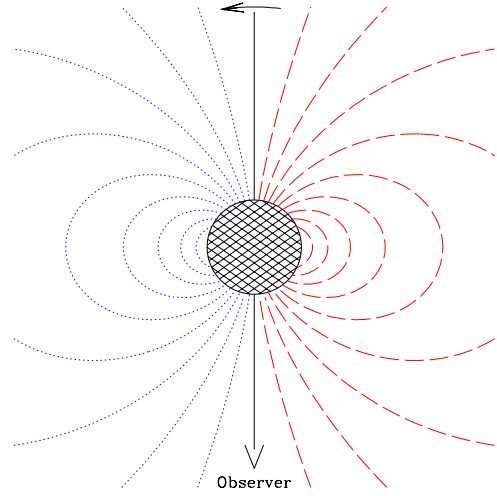}
\caption{Illustration of isovelocity zones for a Keplerian disk.
The hatched region at center is the star, assumed to rotate ccw,
as indicated by the arrow at top.  In this example an observer is
located in the equatorial plane.  Red curves on the right are for
redshifted velocities (dashed), with the smallest loop having the
greatest speed.  Left are for blueshifts (dotted).  Emission at
line center arises from the vertical solid line.
\label{fig2}}
\end{figure}

\subsection{Line Velocity Shifts}

The Sobolev approximation is employed to describe the polarimetric
line profiles.  The Sobolev approach relies on identifying ``isovelocity
zones'', which are the locus of points that share the same Doppler
shift for a distant observer.  The Doppler shift in frequency of
any point in the scattering volume is determined by

\begin{equation}
\Delta\nu_{\rm Z} = -\nu_{\rm ul}\,\frac{\vz}{c} = -\frac{\vz}{\lambda_{\rm ul}},
        \label{eq:nuZ}
\end{equation}

\noindent where $\nu_{\rm ul}$ is the frequency of the line
transition, and $\vz$ is the line-of-sight velocity shift given by

\begin{equation}
\vz = -\vec{v}\cdot\Zhat. 
	\label{eq:vZ}
\end{equation}

\noindent A Keplerian disk follows a velocity profile of the form

\begin{equation}
\vec{v} = \vphi\,\phat = v_0 \, \sqrt{\frac{R_0}{r}}\, \phat,
\end{equation}

\noindent where $R_0$ is the innermost radius of the disk with 
tangential speed $v_0$ at that location.

It is convenient to define dimensionless variables for distances
and velocities.  For the normalized cylindrical radius in the planar
disk, $\varpi = r/R_0$ is used.  The line-of-sight velocity shift
then becomes

\begin{equation}
\vz = v_0\,\varpi^{-1/2}\,\sin i\,\sin \varphi .
\end{equation}

\noindent A normalized velocity is also introduced with $\wz = \vz/(v_0\sin i)$.

The solution for the isovelocity zones in
terms of $\varpi(\wz, \varphi)$ is thus given by

\begin{equation}
\varpi = \frac{\sin^2 \varphi}{\wzz}=\frac{\sin^2 \varphi}{\sin^2 \varphi_0}.
\end{equation}

\noindent This describes a loop path that starts at $R_0$ at an
azimuth $\varphi_0$ on the front side of the disk, where $\sin
\varphi_0 = \wz$, and then ends again at $R_0$ and $\pi-\varphi_0$.
The loop extends to a maximum distance
at azimuth $\varphi=\pi/2$ with a value of $\varpi = 1/\wzz$.  The
system is left-right mirror symmetric, with redshifted velocities
on one side and blueshifted ones on the other.  The convention 
here is that the disk rotates counterclockwise as seen from above
(i.e., ``prograde'').  An example is displayed in Figure~\ref{fig2}.
With the arrow toward the observer as the reference for $\varphi$,
redshifted velocities come from the interval
$0<\varphi<\pi$, and blueshifted ones come from $\pi<\varphi<2\pi$.

\subsection{Line Flux}

For optically thin scattering, the observed flux of scattered light
at a given velocity shift in the line is determined by a volume integral 
along the associated isovelocity loop.  To describe the polarization,
a Stokes vector prescription is adopted, with standard $I, Q, U, V$
notation.  Referring back to Figure~\ref{fig1}, an edge-on disk with
$i=90^\circ$ would yield intensities with $Q>0$ and $U=0$.  In this case
``North'' is chosen to be along the direction of $+\hat{Z}_\ast$, and 
$Q>0$ corresponds to oscillations of the electric vectors of the light
being preferentially parallel to North.  

For unresolved disks the corresponding observed fluxes are $\vec{\cal
F}_\nu = (\FI, \FQ, \FU, 0)$, where it is assumed that the Stokes-V
flux is zero for the problem at hand.

Generally following the notation of Paper~II and using results from
Ignace (2000a) for Keplerian disks, the disk is assumed to have a
power-law surface number density given by

\begin{equation}
\Sigma(\varpi) = \Sigma_0\,\varpi^{-m},
\end{equation}

\noindent for power-law exponent $m$ and inner value $\Sigma_0$
at $\varpi=1$.  Note that in the case of the Be~stars, values of
$m$ range from 3--4 (e.g., Waters 1986; Lee, Osaki, \& Saio 1991;
Porter 1999; Jones, Sigut, \& Porter 2008), and $m=3.5$ will be
adopted in example cases.  Implicit is that the lines of interest
form from the dominant ion.  Temperature and ionization variations
or different disk densities could be included in the model, but the
intent of this contribution is to highlight the line polarization
and Hanle effect, and so a simple power-law density is adopted as
a baseline for the analysis.

The Stokes fluxes for thin lines from this disk are given by

\begin{equation}
\vec{\cal F}_\nu (\wz) = \tau_0\,{\cal F}_0\,\int\,
	\hanvec(\varpi,\varphi)\,\frac{\varpi^{2-m}}{w_\varphi(\varpi)}\,
	\frac{d\varpi}{| \cos \varphi |}.
	\label{eq:flux}
\end{equation}

\noindent The denominator in the integrand arises from the Sobolev
approximation, with $w_\varphi = \varpi^{-1/2}$.  The factor $|\cos
\varphi|$ will more conveniently be written as $\sqrt{1-\wzz\varpi}$
in what follows.  The vector $\hanvec$ represents the ``scattering
factor'' with $\hanvec = (\hI,\hQ,\hU,\hV)$.  It is this factor
that incorporates the dipole scattering of resonance line polarization,
the impact of magnetic fields via the Hanle effect, and the finite
depolarization factor.  Finally,
the scaling factors outside the integral are from equations~(24)
and (25) of Paper~II, with

\begin{equation}
\tau_0 = \frac{\sigma_l\,\Sigma_0\,\lambda_{\rm ul}}{4\pi\,\vrot\,\sin i},
\end{equation}

\noindent where $\sigma_l$ is the frequency integrated line
cross-section, and

\begin{equation}
{\cal F}_0 = \frac{L_{\nu,\ast}}{4\pi\,d^2},
\end{equation}

\noindent where $L_{\nu,\ast}$ is the specific luminosity of the star
and $d$ is the distance to the star from Earth.

\begin{figure}[t]
\plotone{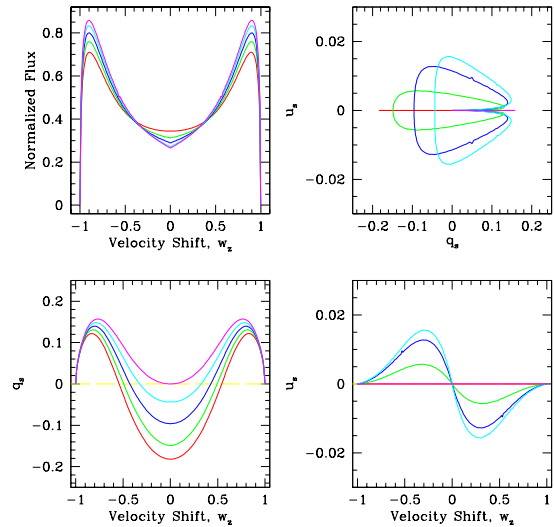}
\caption{Line profiles for a disk with no magnetic field.  Upper
left: Area normalized Stokes-I line profiles for
viewing inclinations of $i=0^\circ$ (red), $i=30^\circ$ (green),
$i=45^\circ$ (dark blue), $i=60^\circ$ (light blue), and $i=90^\circ$
(magenta).  Lower left:  Stokes-Q profiles plotted as the ratio of
the Q-flux to the scattered I-flux as fractional polarization
(not percent).  The yellow dashed line is a guide for zero polarization.
Lower right:  Like the Q-Profiles, Stokes-U profiles shown as
fractional polarizations, which arise solely from the effect of
stellar occultation.  Upper right:  A Q-U diagram across the profile.
\label{fig3}}
\end{figure}

\subsection{Solution for Non-Magnetic Scattering Polarization}	\label{sub:Bzero}

Before progressing to a consideration of the Hanle effect, there
is value in formulating the solution for the line scattering
polarization in the zero field case.  This is because the Hanle
effect primarily {\em modifies} the scattering polarization, meaning
that a good understanding of the non-magnetic case is a necessary
reference for understanding the Hanle effect.  
Many papers have dealt with line formation from
Keplerian disks.  This work draws in particular on the work of Huang
(1961) and Rybicki \& Hummer (1983) for planar disks, and also Ignace
(2000a) for scattering polarization without the Hanle effect.

There are two main cases that are explored here:  the simplistic
case of illumination by a point source and then the more realistic
case involving the effects of the finite size of the illuminating
star.  Analytic and semi-analytic solutions are described in
the Appendices.  The value of the point source approximation is
that it gives context for how finite stellar size effects modify
the line scattering polarization.  The only portion of the point
source approach that is reproduced here from
Appendix~\ref{sec:iso} is the expression for
{\em isotropic} scattering of starlight by a planar disk in
the Sobolev approximation.

For isotropic scattering of light from a point star,
one has $\hanvec = (1, 0, 0, 0)$.  Only
the Stokes-I flux survives in equation~(\ref{eq:flux}).  It is
convenient to introduce a change of variable for computing the
integration of that equation, with $t=\varpi^{-1}$.  Then the flux
becomes

\begin{equation}
\FI = \tau_0\,{\cal F}_0\times 2\,\int_{t_0(w_{\rm z})}^1\,\frac{t^{m-1}}{\sqrt{t-t_0}}\, dt,
\end{equation}

\noindent where $t_0 = \wzz$.  The appearance of the factor of~2
arises because the integration for the front half of the loop (from
$\varphi_0$ to $\varphi=\pi/2$) is the same as for the back half.

Appendix~A details analytic cases for the above case.  The addition
of polarization and finite source effects amount to inserting
multiplicative ``weighting'' functions that modify the above integrand
that describes the emission contribution as a function of location
along a loop.

There are several finite source effects that can be included, such
as finite star depolarization (Cassinelli \etal\
1987), stellar occultation effects (e.g., Fox \& Brown 1991), limb
darkening effects (Brown, Carlaw, \& Cassinelli 1989), and absorption
of starlight by the disk (Ignace 2000a).
Only two of these effects will be considered here:  finite star
depolarization and stellar occultation of the disk.  Limb darkening
could be included; however, limb darkening mainly ``softens'' the
finite star depolarization factor, making it slightly less severe.
Thus the inclusion of limb darkening does not seem to be a pressing
issue for illustrating the Hanle effect in disks.

Absorption by the disk does affect the shape of the Stokes-I profile;
however, in the current treatment it does not influence $\hanvec$
which determines the line scattering polarization.   
Consequently, absorption affects the continuum level
of direct starlight at the line, but not the scattered Stokes fluxes
\FI, \FQ, or \FU.  On the other hand, a photospheric absorption
line would alter the line shapes of the scattered flux profiles;
however, results presented here will be in terms of ratios of
scattered fluxes, $\qs=\FQ/\FI$ and $\us=\FU/\FI$, for which the
influence of a photospheric line will cancel.

These ratios \qs\ and \us\ should be considered as polarimetric
``efficiencies''.  They are not generally what an observer would
actually measure, because the denominator involves only the scattered
light of the Stokes-I flux.  Instead an observer would normally
measure \FQ, \FU, and ${\cal F}_{\rm tot} = {\cal F}_0 + \FI$, the
latter being the sum of the direct starlight (first term) and the
scattered light (second term).  In this case $q_{\rm tot} = \FQ/({\cal
F}_0+\FI) \approx \FQ/{\cal F}_0$, and likewise for the Stokes-U
polarization, assuming that the scattered flux in the line is small
compared to the direct flux by the star itself.  As a result,
$(q_{\rm tot},u_{\rm tot}) = (\qs,\us) \times (\FI/{\cal F}_0)$.
It is worth noting that the \qs\ and \us\ ``efficiency'' profiles
are independent of the line optical depth \taul\ in the thin limit,
since \FI, \FQ, and \FU\ scale linearly with \taul.

Equation~\ref{eq:hIzero} gives the scattering function components
of \hanvec\ in the case of no magnetic field with the star treated
as a point star of illumination.  Modifications to those functions
that allow for scattering by a finite sized and uniformly bright
central star were determined in Paper~III.  For the case of zero
magnetic field, the scattering function is shown explicitly with
factors arising from the finite stellar size.  These factors also
appear in the scattering function with the Hanle effect in the same
positions as without the Hanle effect.  The vector components of
\hanvec\ are:

\begin{equation}
\vec{h} = \left\{ \begin{array}{lcl} 
	h_I & = & \tilde{W}+\frac{3}{8}E_1\,\mustar\,\left[\frac{1}{3}\,
		\left(1-3\cos^2 i
		\right) \right. \\
            &   & \left. + \sin^2 i\,\cos 2\varphi \right] \\
	h_Q & = & \frac{3}{8}E_1\,\mustar\,\left[\sin^2 i \right. \\
	    &   & \left. -\left(1+\cos^2 i \right)\,\cos 2\varphi \right] \\ 
	h_U & = & \frac{3}{8}E_1\,\mustar\,\cos i\,\sin 2\varphi, 
	\end{array} \right.	\label{eq:finite}
\end{equation}

\noindent where 

\begin{equation}
\tilde{W} = 4\varpi^2\,W(\varpi),
\end{equation}

\noindent with the dilution factor $W(\varpi)=0.5\,
(1-\mustar)$, and

\begin{equation}
\mustar = \sqrt{1-\varpi^{-2}}
\end{equation}

\noindent is the finite depolarization factor when there is no limb
darkening.  The factors $\tilde{W}$ and $\mustar$ represent corrections
to the scattering function that account for the effects of finite
star size in relation to the incident radiation field at a scattering
point around the star.  

Scattering by resonance lines can be approximated as part dipole
and part isotropic.  The parameter $E_1$ is a factor with values
from 0 to 1 that represents the extent to which a resonance line
scatters like a dipole radiator (see Chandrasekhar 1960).  Only the
dipole portion of the scattered light contributes to observable
polarization.  In the solar literature, it is more common to use
the notation $W_2$ for the fraction of scattered light that is
dipole-like (e.g., Stenflo 1978).

Accounting for stellar occultation breaks the back-front symmetry
of the isovelocity loops.  The front loop still has lower and upper
limits of $t_0$ and 1, but the back half has limits of $t_0$ and
$t_\ast\le 1$.  Here $t_\ast$ is associated with the minimum value of
$\varpi = \varpi_\ast$ at the back-projected limb of the star
corresponding to a maximum azimuth of $\varphi=\varphi_\ast$.  An
expression for $\varpi_\ast$ is given in Ignace (2000a), 
expressed here as 

\begin{equation}
t_\ast = \sqrt{1-\sin^2 i\cos^2\varphi_\ast} = \sqrt{\cos^2 i-\wzz\sin^2i\,
	/t_\ast}.
\end{equation}

\noindent This expression provides a cubic relation in $t_\ast$ as
a function of fixed viewing inclination $i$ and velocity shift \wz.

Numerical results for polarized profiles without the Hanle effect
are shown in Figure~\ref{fig3}.  This Figure and the ones that
will follow use $m=3.5$ and $E_1=0.5$.  Figure~\ref{fig3} should
be compared to Figure~\ref{figApp1} for the point star case.  For
point star illumination, the maximum polarization occurs at the
line wings and $\FU\equiv 0$ by symmetry.  This changes dramatically
when the effects of the finite stellar size are included.

In Figure~\ref{fig3}, each panel has 5 curves color-coded for
different viewing inclinations:  $\sin^2 i= 0$ is red, $\sin^2 i=
0.25$ is green, $\sin^2 i= 0.5$ is deep blue, $\sin^2 i= 0.75$ is
light blue, and $\sin^2 i= 1.0$ is magenta.  The values correspond
to viewing inclination angles of $i=0^\circ, 30^\circ, 45^\circ,
60^\circ,$ and $90^\circ$, respectively.  The upper left panel
displays \FI\ profiles that are normalized to have unit area.  In
the lower right panel for the \qs\ polarization, the finite
depolarization factor shifts peak polarization to lower velocity
shifts as compared to the point star case, and \qs\ remains symmetric
about line center.  Stellar occultation leads to the survival of a
small \us\ polarization as shown in the lower right panel.  The
upper right panel is a \qs-\us\ plot across the respective line
profiles.  Note that the scales in \qs\ and \us\ differ by about
an order of magnitude.  These small loops signify position angle
rotations across the observed polarized line profiles, with
polarization position angle $\psi_P$ given by $\tan 2\psi_P =
\us/\qs$.

We next turn our attention to the Hanle effect.  The results of this section
will prove valuable in following how the Hanle effect modifies the run
of polarization across the line profile.

\section{THE HANLE EFFECT IN KEPLERIAN DISKS}	\label{sec:hanle}

The Hanle effect applies to resonance line scattering
and can be interpreted in
semi-classical terms as a precession of an oscillating emitter that occurs
over the radiative lifetime of the line emission.  An angular quantity can
be defined to represent the effective amount of precession, as given by

\begin{equation}
\tan \alt = \frac{2\,g_L\,\omega_L}{A_{\rm ul}} = \frac{B}{\BHan},
        \label{eq:tanalt}
\end{equation}

\noindent where $\omega_L$ is the angular Larmor frequency, $A_{\rm
ul}$ is the Einstein radiative rate for the transition of interest,
$g_{\rm L}$ is the Land\'{e} factor of the upper level, $B$ is the
magnetic field strength in Gauss, and \BHan\ is the Hanle field
sensitivity defined as

\begin{equation}
\BHan = 56.9\,\frac{A_9}{g_{\rm L}}~{\rm Gauss},
\end{equation}

\noindent with $A_9$ the radiative rate normalized to $10^9$~Hz.
The Hanle effect manifests itself in the scattering physics with
the appearance of cosine and sine functions of the angle \alt\ 
(and a related quantity $\alpha_1=0.5\alt$) within
the scattering functions \hanvec.  Note that with no magnetic field,
$B=0$ and $\alt=0$.  The limit of a high Larmor frequency yields
$\alt \approx \pi/2$.  In this latter case, the Hanle effect is
said to be ``saturated''.

There are a few rules of thumb to help understand the Hanle effect
in polarized lines.  First, modifications to the polarized profile
will depend on the orientation of the field in relation to both the
direction of incoming radiation and outgoing radiation (i.e., the
observer).  Second there is no Hanle effect when the incoming
radiation field is symmetric about the field direction.  These first
two ``rules'' indicate that there is no Hanle effect for radial
magnetic fields (assuming a spherically symmetric star and
no diffuse radiation) and that
only non-radial components of the field contribute to modifying the
polarization.  However, the third point is that in spite of this,
the angle \alt\ is sensitive to the {\em total} field strength:  both
radial and non-radial components, at the site of scattering.
Fourth, and finally, the Hanle effect is sensitive only to the
field topology in the saturated limit, not the field strength.

With these rules of thumb, consideration of the Hanle effect in
three particular cases follow:  a purely axial field, a purely
toroidal field, and last a scenario involving fields as arising
from the operation of MRI in disks.

\subsection{Axial Magnetic Field}

In this case an axial magnetic field is envisioned as threading the
disk perpendicular to the equatorial plane, hence $\vec{B} =
B_{Z_\ast}(\varpi)\,\hat{Z_\ast}$.  As for the field distribution
throughout the disk, the following is adopted as a conservative
limiting case:

\begin{equation}
B_{Z_\ast} = B_0 \, \varpi^{-1},
\end{equation}

\noindent and so the Hanle angle \alt\ is determined by

\begin{equation}
\tan \alt = \frac{B_0}{\BHan}\,\varpi^{-1}\equiv b_0\,t .
	\label{eq:tanalt}
\end{equation}

\noindent where $b_0=B_0/\BHan$ for $B_0$ a field strength at the
inner disk radius.  It is important to note that if $B_0\ll \BHan$,
then the Hanle effect is weak everywhere in the disk.  If $B_0\gg
\BHan$, then the inner disk will be in the saturated limit; however,
there will be a transition to a weak Hanle effect at some radius
in disk, characteristically where $\varpi = B_0/\BHan$.  As a general
rule, for a given magnetized disk, different lines will have different
values of \BHan, and thus will be sensitive to different aspects
of the field at different locations in the disk.

\begin{figure}[t]
\plotone{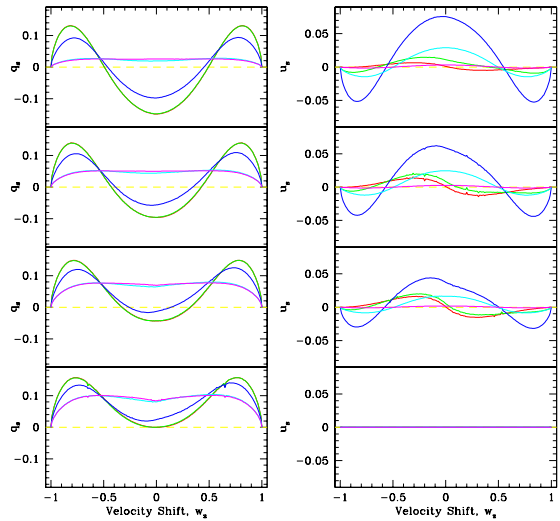}
\caption{The Hanle effect for a disk threaded by a purely axial
magnetic field.  Left is for \qs\ and right is for \us.  Profiles
are shown for different viewing inclinations with $i=30^\circ$
at top, then $i=45^\circ$, $i=60^\circ$, and $i=90^\circ$ at
bottom.  The colors are for different field strengths with
$b_0=0.01$ (red), 0.1 (green),  1.0 (dark blue), 10 (light blue),
and 100 (magenta).
\label{fig4}}
\end{figure}

\begin{figure}[t]
\plotone{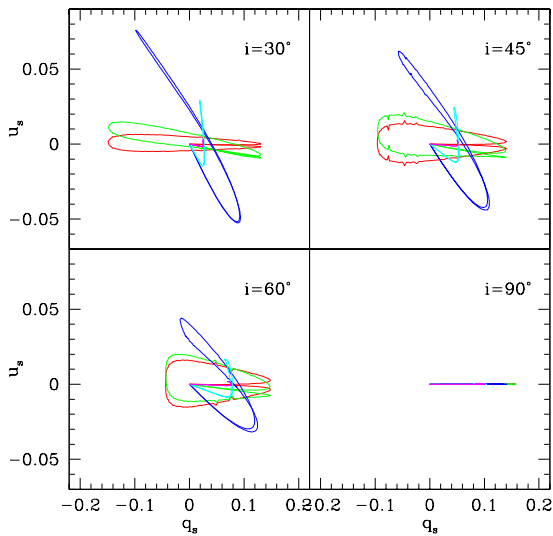}
\caption{
A Q-U diagram for the polarized profiles shown in Figure~\ref{fig4}
with the same color scheme in relation to results for different values
of $b_0$.  The strongly ``saturated'' case of $b_0=100$ is present
in magneta.  This case has a modest range in \qs\ from 0 to
about 0.1, but with $\us\approx 0$ everywhere, the profile is only
a short horizontal
line in the Q-U plane and difficult to see in this figure.  For
the edge-on case, $\us\equiv 0$ and all profiles degenerate to
horizontal lines from 0 to a maximum value of \qs\ that depends
on $b_0$.
\label{fig5}}
\end{figure}

Expressions for \hanvec\ in the point star limit are given in
Appendix~\ref{app:axial}.  Revisions to those expressions for finite
star depolarization and stellar occultation is the same as in
equation~(\ref{eq:finite}).  Results
are shown in Figure~\ref{fig4}.  Profiles of \qs\ and \us\ are
displayed at left and right, respectively.  Panels from top to
bottom are now for different viewing inclinations of $\sin^2 i = 0.25,
0.5, 0.75,$ and 1.0.  Different colors are for different values of
$b_0$ with $b_0=0.01$ (red), 0.1 (green), 1.0 (dark blue), 10
(light blue), and 100 (violet).

The Stokes-Q profiles are symmetric about line center, whereas the
U~profiles evolve from weak and anti-symmetric, owing to stellar
occultation, to strong and mostly symmetric, to a null profile at
the strongly saturated limit.  In the point star case, the saturated
limit yields $\us=0$ and $\qs=constant$ for all velocity shifts.
With finite source effects, \qs\ is mostly flat throughout the
central portion of the profile, but goes to zero in the wings.
In the presence
of limb darkening, the \qs\ profile would become more flattened
toward the wings.

Figure~\ref{fig5} displays \qs-\us\ curves 
across the line profile.  Note that the two axes have different
scales:  variations in \us\ are smaller than for \qs.  The colored
curves are the same cases as shown in Figure~\ref{fig4}.  Viewing
inclinations are indicated.  For the edge-on
case of $i=90^\circ$, the curves are all degenerate with $\us=0$.
The red loop is for a very weak field and is seen to be top-bottom
symmetric in this space.  As the field is increased, \qs\ and \us\
profiles become individually
more nearly symmetric.  In the \qs-\us\ space, this
results in curves with only small loops, for example the modest
field case with the dark blue curve.  As the disk enters the saturated
limit, \us\ drops to zero, \qs\ becomes somewhat flat-topped in
appearance (again, except at the line wings), which degenerates
mainly to a point in the \qs-\us\ space for most velocity shifts,
with extension to zero polarization only for the line wings.

\subsection{Toroidal Magnetic Field}

A completely azimuthal field configuration has also been
considered.  As in
the axial field case, the toroidal field strength is assumed to
decrease inversely proportional to $\varpi$, and so
equation~(\ref{eq:tanalt}) remains valid.  This means
that differences in the model polarized profiles between the axial
and toroidal field configurations
arise because the Hanle effect is sensitive to the vector field
orientation.

\begin{figure}[t]
\plotone{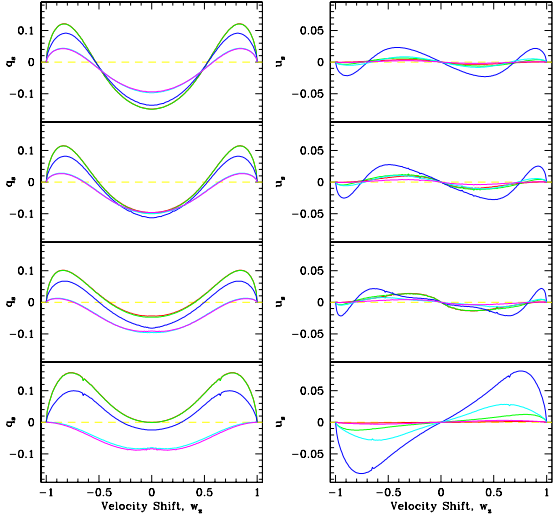}
\caption{Similar to Figure~\ref{fig4} but now for a toroidal field
instead of an axial one.  From top to bottom, panels are for
the same viewing inclinations, and the color scheme is for the same
values of $b_0$.  The \qs\ profiles are symmetric although
differing in details from the axial field case.  More striking is
that the \us\ profile is decidedly antisymmetric for the $B_\varphi$
case, whereas profiles tend toward being symmetric in the axial
field case.
\label{fig6}}
\end{figure}

As with previous considerations, some analytic results are derivable
in the point star limit, and these are detailed in
Appendix~\ref{app:toroidal} for a disk with a toroidal field.  With
a toroidal field, the geometry associated with determining the Hanle
effect is more complex than in axial field case.  Geometrical
relationships between the various angles defining the scattering
problem with a toroidal field in a disk were given in the Appendix
of Paper~II and will not be repeated here.  Results for the calculation
of polarized profiles, including the effects of the star's finite
extent, are shown in Figure~\ref{fig6}.  This Figure is presented
in the same manner as Figure~\ref{fig4} for the axial field case:
\qs\ is at left and \us\ at right; the panels are for different
values of $\sin^2 i$; the colors are for different field
strengths characterized by $b_0$.

\begin{figure}[t]
\plotone{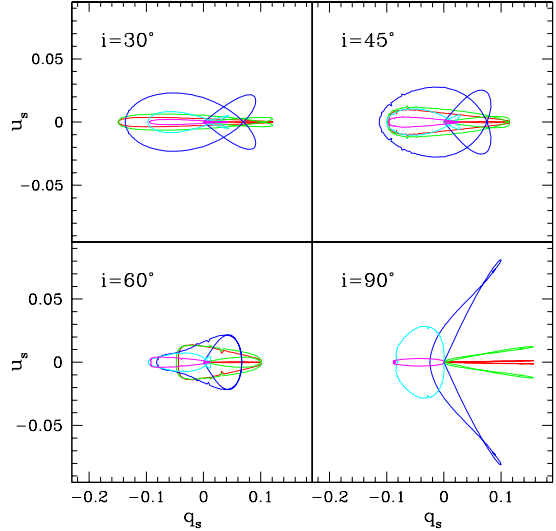}
\caption{Similar to Figure~\ref{fig5} but now for a toroidal
field instead of an axial one.  The combination of antisymmetric
\us\ with symmetric \qs\ leads to Q-U loops that are top-bottom
symmetric in this space.  Note that the axis scale is different
for \us\ as compared to \qs.
\label{fig7}}
\end{figure}

Note the marked differences in the line polarization between the
toroidal case and that of an axial field.  The Stokes-U profiles
are always antisymmetric.  Like the axial case, \qs\ profiles are
symmetric, but the behavior is quite different.  For example the
edge-on disk case at lower left indicates that the profile polarization
completely changes sign, from positive definite at every velocity
shift when $B=0$ to everywhere negative in the saturated limit.
This implies a $90^\circ$ rotation in position angle between these
two limiting cases for the entire profile.  For intermediate field
strengths, there is a position angle rotation across the profile
that occurs at different velocity shifts.  In the axial field case,
there is never a position angle rotation for an edge-on disk.  For
the \qs-\us\ shapes, differences in the polarizations are especially
clear, with results for the toroidal field in Figure~\ref{fig7} to
be compared against those for an axial field in Figure~\ref{fig5}.

What is the source of these differences between the axial field and
the toroidal one?  The field orientation with respect to the viewer
is clearly the key to the interpretation.  For the axial case, the
field at every point in the disk has some component of the magnetic
vector directed toward the observer (if seen at $i<90^\circ$) or
away from the observer (if seen at $i>90^\circ$).  Thus, an axial
field leads to a net projected magnetic flux of one sign or the
other, which is true for every isovelocity loop.  The toroidal field
is manifestly different, since one side of the disk has components
toward the observer and the other side has them away.  For an
axisymmetric toroidal field, the projected net magnetic flux is
identically zero for the entire disk, but is oppositely signed in
isovelocity loops for blueshifted velocities versus redshifted ones.
In terms of the semi-classical precession description, flipping the
field by $180^\circ$ amounts to a precession in the opposite
direction.  In the axial field case, the precession of the radiating
oscillator is uniform -- entirely clockwise (cw) or counterclockwise
(ccw).  Both cw and ccw precessions occur for a toroidal field
configuration, with one precession occurring in half the line
profile, and the opposite for the other half.

\begin{figure}[t]
\plotone{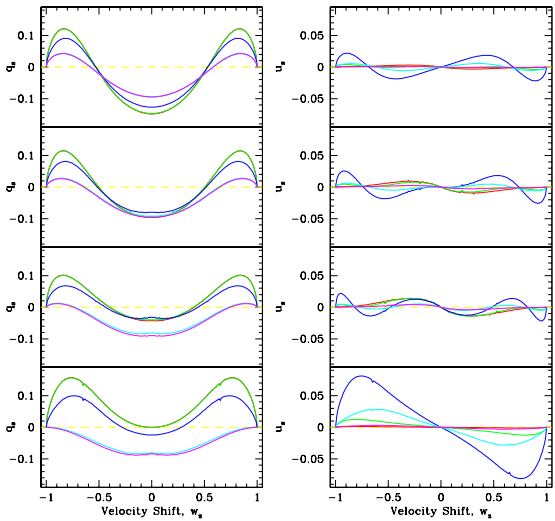}
\caption{The same cases shown in Figure~\ref{fig6} except
now for a toroidal field with the opposite sense of rotation.
Note how the \us\ profiles are essentially the reverse of
those shown in Figure~\ref{fig6}.
\label{fig8}}
\end{figure}

To illustrate this effect, Figure~\ref{fig8} shows line profile
results for a toroidal field that now goes in the other direction
as compared to Figure~\ref{fig6}.  The \qs\ profiles are nearly the
same, but the \us\ profiles are nearly mirror images of one another.
slight differences are a result of the stellar occultation,
which does {\em not} flip when the field orientation is reversed.

\subsection{Magneto-Rotational Instability}

A large literature has developed over the last twenty years in
relation to the magnetorotational instability (MRI; Balbus \& Hawley
1991, 1998).  There is an interesting history about this effect
(e.g., Balbus 2003).  The instability can be illustrated through
an analogy to two masses in a differentially rotating disk that are
slightly offset from each other along a radius.  These two masses
are connected by a spring, to represent the effect of an axial
magnetic field.  The end result is that the two masses on different
orbits seek to increase their displacement from one another.  Coupling
by the spring leads to a runaway situation ensues.

Key for the context of magnetic diagnostics is how this 
instability impacts the field topology and strength throughout the
disk.  The mass-spring analogy above has built into it the assumption
of flux freezing.  The separation of the masses along with the
differential rotation would appear to evolve the axial field through
the disk into a toroidal one.  Different researchers have studied
the operation of the MRI in accretion disks through both semi-analytic
work and numerical simulations (e.g., Balbus \& Hawley 1991, 1992;
Hawley \& Balbus 1991, 1992; Hawley, Gammie, \& Balbus 1995; Hawley
2000; Fromang \& Stone 2009; Lesur \& Longaretti 2009; Maheswaran
\& Cassinelli 2009).

The goals of the MRI simulations are to understand better the physics
of angular momentum transport through disks and disk structure.
However, this paper seeks better insight into whether and how disk
magnetism might be directly detected, with a focus on the Zeeman
and Hanle effects for spectral lines.  Ignace \& Gayley (2008)
reported on a simplistic calculation of the Zeeman effect and the
Hanle effect for a Keplerian disk with a purely toroidal field.
Here, application of the Hanle effect to Keplerian disks has been
developed in a more complete way.  But before discussing its
application to the MRI scenario, it is worth commenting on the
conclusions of Ignace \& Gayley for the use of the Zeeman effect.

The MRI leads to a field topology that consists of a toroidal
component and a randomized component.  The toroidal components
exists in mixed polarity, by which it is meant that some sectors
run cw and others run ccw.  For the weak
longitudinal Zeeman effect, the Stokes-V flux scales with the net
magnetic flux associated with a spatial resolution element (e.g.,
Mathys 2002).  With bulk motions frequency (or wavelength or velocity)
resolution ultimately maps to geometrical zones at the source (an
example of this involving the Sobolev approximation was detailed
by Gayley \& Ignace 2010 for spherical winds).  The key point is
that if both the randomized field and the polarity of the toroidal
component changes on small scales, then the flux of circularly
polarized light arising from the Zeeman effect will be strongly
suppressed owing to little net magnetic flux.

The issue of variations of the field on ``small scales'' must be
evaluated with care.  In the Sobolev approximation for purely
Keplerian rotation, we have seen that the isovelocity zones are
``loop'' structures.  These loops extend out to large radius for
low velocity shifts, but they can be quite small at high velocity
shifts.  The highest velocity shifts degenerate to a pair of points
for either limb of the star at the projected stellar equator!  More
importantly, every isovelocity zone intersects with the innermost
radius of the disk.  (If the disk extended down to the star, this
would be the photospheric radius.)  A steep power-law density ensures
that the bulk of emission or scattered light comes from inner disk
radii.  As a result, the Zeeman effect would be most sensitive to
a magnetic field at these locations, and thus ``small scales''
refers to turnovers in the field direction that are small compared
to the inner radius of the disk.  Consequently, the detection of
the Zeeman effect from a disk with a given spectral resolution and
density distribution constrains the characteristic spatial wavelengths
at which the field turns over with radius from the star, azimuth
around the star, and vertical height through the disk.

A detection of the Zeeman effect in a disk has previously been
reported by Donati \etal\ (2005) in the case of FU~Ori.  This is
thought to be an accretion disk with a disk wind, as evidenced by
some lines showing P~Cygni absorption.  However, the circular
polarization profile is argued as being associated with the innermost
region of the rotating disk where the field is of kilogauss strengths.
The detection implies a net magnetic flux per spectral resolution
element, and thus sets limits on the turnover (or ``tangledness'')
length scale for the disk field, if indeed the MRI is operating in
this case.

It is interesting to consider signals that could result with the
Hanle effect.  The Hanle effect can operate in regions where magnetic
fields are ``tangled'' or randomized.  This means that spatial
averages of $\langle \vec{B} \rangle$ tend toward zero although
$\langle B^2 \rangle$ does not, such as is the case for the MRI
mechanism.  An extensive literature exists for the Hanle effect
with random fields in applications to solar studies (e.g.,
Frisch \etal\ 2009, and references therein).
Here I simply want to outline some of the limiting behavior in
applications to disks.

Consider a Keplerian disk that contains everywhere a truly randomized
magnetic field.  If the field is weak at all locations, meaning
that $B \ll \BHan$, then of course a \qs\ line profile results as
in the case of no Hanle effect.  But if the field is strong, such
that the denser regions of the disk are largely saturated, then the
behavior is much different.  With different field orientations, one
expects that \us\ will yield a null profile, by symmetry considerations.
However, the \qs\ profile is different.  

As a specific example, consider the resulting \FQ\ line from an
edge-on disk.  Without a field, the polarization at line center
would be zero, owing to forward scattering of unpolarized starlight.
With a randomized field, the polarization will still tend toward
zero at line center.  In the point star limit, some net polarization
is expected to survive in the line wings.  This polarization will
be significantly reduced in comparison to the zero field case.  If
$E_1$ were unity, the polarization at the line wings would be 100\%,
since the scattering geometry is $90^\circ$.  In analogy with
considerations of scattering polarization off the solar limb, a
reduction in polarization by a factor of 5 for isotropically
distributed fields should be expected (see Stenflo 1982).

Now consider the introduction of a sustained
toroidal field component.  Of course toroidal fields were
considered in the previous section.  Now however, the toroidal
field has polarity flips within the disk -- $B_\varphi$ is
alternately cw or ccw at essentially random
points within the isovelocity zones.  What this means is that there
is a sign change in the direction of Larmor precession in the
classical picture of a harmonic oscillator.  The effect of this is
to drive the \us\ signal to zero faster than if the toroidal field had one
sense of polarity.  In the saturated limit, the \qs\ profile is
{\em unchanged}, because the surviving polarized signal does not
depend on polarity at all.  This is quite different from the Zeeman
effect that is sensitive to the net magnetic flux.  For the Zeeman
effect, the circular polarization will be suppressed when $B_\varphi$
switches polarity on small spatial scales.

\section{CONCLUSIONS}	\label{sec:conc}

Polarized line profile shapes from magnetized Keplerian disks have
been calculated under a number of simplifying assumptions:  the
disk is geometrically thin; the scattering lines are optically thin;
primarily simple fields were considered (axial or toroidal); and
no account was taken of photospheric absorption lines.  On the other
hand, the model profiles do include finite source depolarization
and the effects of stellar occultation.  The presentation of
results focused on the polarimetric ``efficiencies'' \qs\ and \us\
with a description of how to identify the occurrence of the Hanle
effect in scattering lines from disks.  In addition, a discussion
was presented for the Hanle effect from a magnetized disk in which
the MRI mechanism is operating.

One of the main conclusions from this work is that axial
and toroidal fields in disks are easily distinguishable through an
analysis of \qs-\us\ figures for the run of Stokes polarizations
across line profiles.  Although a \us\ profile does exist even
without the Hanle effect, owing to stellar occultation, its amplitude
is quite small.  The strongest \us\ profiles result when much of
the inner disk, where most of the scattered light is produced, has
values of $B/\BHan$ of order a few.  If the inner disk is mostly
in the saturated limit of the Hanle effect, the \us\ profile becomes
a null profile for both axial and toroidal fields.

One should bear in mind that the polarized efficiencies are upper
limits to the polarizations that would actually be measured,
since the efficiencies are with reference to the scattered
light only and do not take account of the direct starlight from
the system.  Since this starlight is expected to be largely
unpolarized, its contribution acts to ``dilute'' the polarization
substantially below the efficiency levels reported here.
For example, if the line is relatively weak at 20\% of the continuum
level, then the expected measured polarizations would have fractional
values at about 1/5 of the efficiency values, resulting in line
polarizations of around 1\%--2\% for \qs\ and 0.5\% or less for \us.

As percent polarizations, such values would appear to be easily
measurable, but in practice there are several challenges.  First,
a spectral resolution yielding several points across the polarized
profile is needed.  Circumstellar disks have rotation speeds of
order 500~km~s$^{-1}$.  A requisite velocity resolution of
perhaps 50~km~s$^{-1}$ would then be needed, 
implying  a resolving power of
$\lambda/\Delta \lambda \approx 6000$.  Harrington \& Kuhn (2009a)
have demonstrated that such resolving powers can be achieved in
spectropolarimetry; however, the next requirement is that of finding
suitable scattering lines.

The very interesting effects seen in the sample of Harrington \&
Kuhn (2009a) exploits a relatively new effect of enhanced polarization
that is coincident with regions of higher line absorption, a
consequence of optical pumping effects.  For the Hanle effect, or
even for non-magnetic resonance scattering, lines that are
predominantly scattering are needed.  For hot star disks, such as
the disks of Be~stars, resonance scattering lines are generally to
be found at UV wavelengths.  This requires space-borne spectropolarimeters.
Although the Wisconsin Ultraviolet Photo-Polarimeter Experiment
(WUPPE) obtained exciting new results from UV polarimetry, its
resolving power was only about 200 (Nordsieck \etal\ 1994).  A new
UV spectropolarimeter called the Far-Ultraviolet SpectroPolarimeter
(FUSP) is a sounding rocket payload that will have a resolving power
of about 1800 (Nordsieck 1998).  Although possibly two low for
circumstellar disks, it will be suitable for studying scattering
line polarizations from high velocity stellar wind sources.

As a matter of practical analysis, how should spectropolarimetric
data best be managed to measure a Hanle effect?  Plotting the
velocity shifted polarizations in the \qs-\us\ space appears to be
most promising.  The sequencing of the analysis for scattering lines
from disk sources might proceed as follows:

\begin{itemize}

\item Subtract off the continuum polarization in the vicinity of the line
of interest.  This will counter the effects of both interstellar polarization
and any other broadband source polarization, such as may arise from
Thomson scattering in the disk.

\item Plot the Q and U line {\em fluxes}, not fractional or percent
polarizations that would be derived through normalization by the
total I-flux.  The reason for plotting polarized fluxes is that
many common UV resonance lines are doublets (e.g., N{\sc v}, Si{\sc
iv}, and C{\sc iv} UV resonance doublets).  The shorter wavelength
component (``blue'') has $E_1=0.5$, and the long wavelength component
(``red'') has $E_1=0$ (e.g., Tab.~1 of Paper~II).  If the lines are
thin, then the polarized flux from the red line will not be influenced
by the blue one.  However, if the lines are sufficiently closely
spaced, then \FI\ will be a blend, and normalization by that blend
would artificially skew the Q-U figure shape, making interpretation
more difficult.  If the doublet components are well separated, then
relative polarizations could also be used in what follows.

\item Determine whether the resultant figure for the line polarization
shows any axis of symmetry.  If so, then either (a) there is no
Hanle effect, or (b) there is a Hanle effect with a dominant toroidal
component.  For an axis of symmetry, a rotation of the figure from
observer Q-U axes to a source set of axes Q'-U' could be accomplished
with a Mueller rotation matrix.  After rectifying the figure in this
way, the relative amplitude of \FU\ should be compared to \FQ.
If $\FU \ll \FQ$, then there is likely little Hanle effect or
the disk is in the saturated limit of the Hanle effect.  If $\FU \sim
\FQ$, then a Hanle effect is required with $b_0 \approx$ few in the
disk where the bulk of scattered light is produced.  This means 
$B \approx \BHan$.  

\item If there is no symmetry axis to the Q-U figure, then a Hanle effect
involving an axial field is most likely the culprit.  Recall that
a radial field could be present, but this would give no Hanle effect
by itself.

\item For identifying a field distribution arising from MRI, things
are more complicated.  If the Hanle effect is in operation, then
the Stokes-U flux is likely driven to zero, even if $b_0 \sim 1$.
A net \FQ\ profile should survive; however, it may not be symmetric.
Using an oversimplification, each isovelocity zone can be considered
to have a different effective $b_0$ value.  This is already the
case in the pure axial or toroidal field case that is axisymmetric,
but the variation of the effective $b_0$ with velocity shift is
monotonic (in a flux-weighted sense).  With MRI, one expects
non-monotonicity from fluctuations of the field strength.  
This amounts to the introduction of amplitude fluctuations across
the polarized line. 

\end{itemize}

There are complications to the above approach.  For example, in the
case of the Be~star disks, there is substantial evidence for one-armed
spiral density wave patterns that make the disk density non-axisymmetric
(e.g., Okazaki 1997; Papaloizou \& Savonije 2006).  For thin Thomson
scattering, there is little observational consequence of this effect
which is mainly a redistribution of disk scatterers in a point
antisymmetric pattern (e.g., Ignace 2000b).  However,
with resolved line profiles, the non-axisymmetric density distribution
will produce asymmetry in the \qs\ profile and will be a new source
of net \us\ signal, also asymmetric.  Certainly it will be important
to model the \FI\ profile self-consistently along with the polarized
profiles.

Note that there may be some concern about the choice adopted
for the radial distribution of the field strength through the disk.
The choice of $|\vec{B}| \propto \varpi^{-1}$ is perhaps the most
shallow distribution that one could expect.  A steeper gradient of
the field strength will restrict the operation of the Hanle effect
to a more restricted range of radii in the disk.  If the interval
in radius where $b_0 \gtrsim 1$ becomes narrow in relation to where
most of the scattered light is produced, then the Hanle effect will
be irrelevant for the observed polarization.  

What are the next steps in formulating better diagnostics of the
density and magnetic field structure in disks?  Most of work
considered in this series of papers has focused on optically thin
lines in an attempt to gain a better understanding of how the Hanle
effect influences line polarizations that form in circumstellar
media.  In Paper~IV the issue of optical depth effects for P~Cygni
lines from stellar winds were considered by treating regions of
high optical as contributing no polarization at all.  Although
simplistic, insight was gained into how optical depth effects provide
an additional spatial ``filter'' in terms of where bulk of line
polarization will be produced.  Naturally, rigorous radiative
transfer and a more realistic disk model (i.e., not simply planar)
is badly needed to extend the considerations of this paper.  Even
with thin line scattering, it would be useful to explore how
non-axisymmetric disk models, such as the one-armed spiral density
wave pattern for Be~disks, will modify the line polarization.
Moreover, winds driven off magnetized disks (e.g., K\"{o}nigl 1989;
Proga, Stone, \& Drew 1998; Proga, Stone, \& Kallman 2000) have
been ignored entirely in this work.  These are all new calculations
that will need to be considered in the future.

\acknowledgements

The author is indebted to the anonymous referee whose comments have
improved this manuscript.  Appreciation is expressed to Ken Gayley,
Joe Cassinelli, and Gary Henson for engaging conversations about
line polarizations and the Zeeman and Hanle effects.  This work was
supported by a grant from the National Science Foundation (AST-0807664).

\appendix

\section*{APPENDIX:  SPECIAL CASES FOR THE LINE PROFILE SHAPES
	}

In these Appendices, a number of instructional cases are considered
for the polarized line profiles shapes from a Keplerian disk when
the illuminating star is treated as a point source.  This means
that both stellar occultation and the finite star depolarization
factor are ignored.  A consequence of this approximation is that a
non-zero \us\ profile can {\em only} result from the Hanle effect.
Before considering polarized line profiles, Stokes-I profile shapes
are derived for the case of isotropic scattering.  These solutions
form the base emissivity function from which the polarized lines
are constructed.

\begin{figure}[t]
\caption{Analytic line profile shapes from isotropic scattering for
the density exponent parameter $m=3$ at left and $m=4$ at right.
The flux profiles are shown at top and \qs\ profiles at bottom.
The different curves are for different viewing inclinations,
with $\sin^2 i = 0.0, 0.25, 0.5, 0.75$ and 1.0.  In each case
$E_1=0.5$.  The flux profiles are normalized to the total
line flux in the {\em isotropic} case.  The normalized isotropic
line is plotted in light blue in the upper panels.
\label{figApp1}}
\plottwo{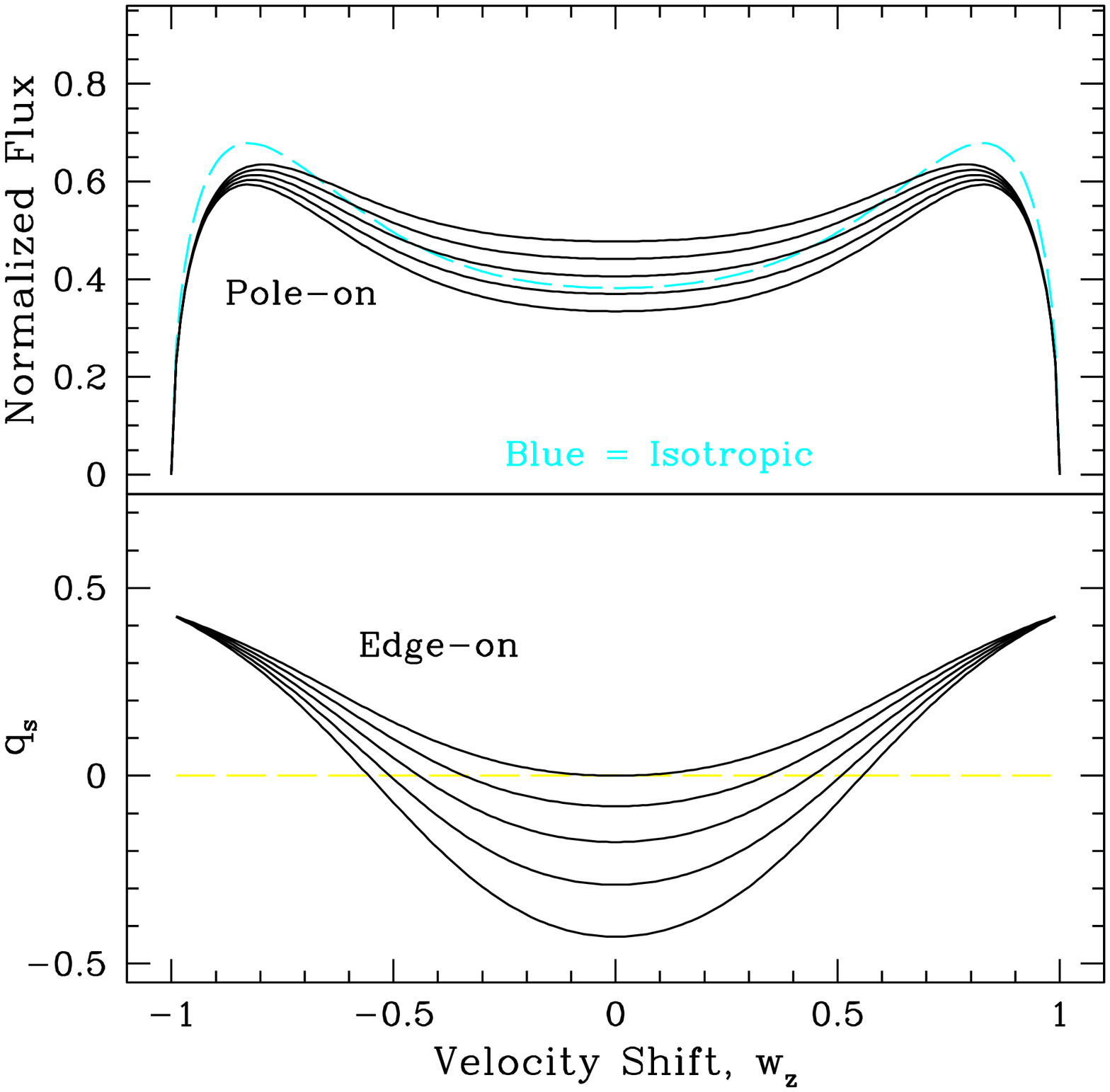}{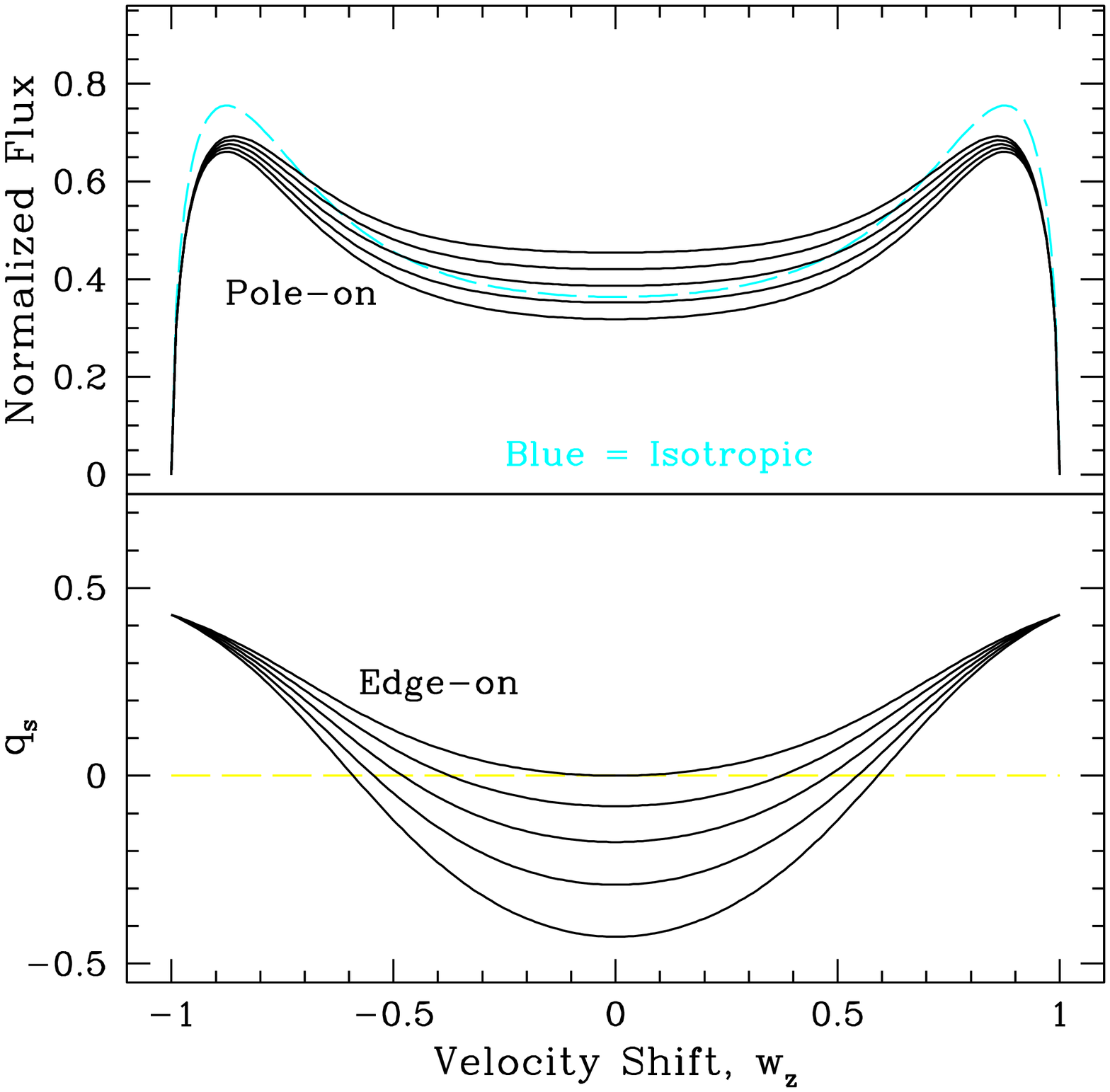}
\end{figure}

\section{The Case of Isotropic Scattering}	\label{sec:iso}

Isotropic scattering corresponds to $E_1=0$, and it means there is
no polarization from resonance line scattering.  Of course, that
also means there is no Hanle effect, regardless of the field strength.

Even though there is no Hanle effect, the isotropic case is useful
to explore as a reference for the production of the Stokes-I line
shape.   The integrand for the line emission as a function of
velocity shift in the observed line represents the contributions
by the disk density and the Sobolev effect for the profile shape.
Allowing for $E_1 \neq 0$ and the Hanle effect simply represents new weighting
functions for non-isotropic scattering that multiply the integrand
from the isotropic case.

The flux of line emission at normalized Doppler shift $\wz$ is

\begin{equation}
\fiso = \taul\,{\cal F}_0\times 2\int_{t_0}^1\,\frac{t^{\rm m-1}}
	{\sqrt{t-t_0}} \, dt,
\end{equation}

\noindent where the factor of 2 arises from the back-front symmetry
of the integration along the isovelocity zone. As a reminder,
$t=\varpi^{-1}$ and $t_0=\wzz$.  

Again the preceding expression is only valid in the point star
approximation.  The power law exponent $m$ is from the surface
density distribution that is assumed to be a power law of the form
$\Sigma = \Sigma_0\, \varpi^{\rm -m}$.
This formulation leads to symmetric double-peaked line profile
shapes for $m>2$.  Larger values of $m$ result in profiles that
have more pronounced double-horns at greater velocity shifts from
line center.  

For integer values of $m$, the integral is analytic, and solutions
for $m=3$ and $m=4$ are given here by way of example.  For $m=3$
the result is

\begin{equation}
\fiso(3) = \taul\,{\cal F}_0\times 2\int_{t_0}^1\,\frac{t^2}
        {\sqrt{t-t_0}} \, dt, = \taul\,{\cal F}_0\,\times 
	\frac{4}{15}\,
	\left(3+4\wzz+8w_{\rm z}^4\right)
	\,\sqrt{1-\wzz},
\end{equation}

\noindent and for $m=4$,

\begin{equation}
\fiso(4) = \taul\,{\cal F}_0\times 2\int_{t_0}^1\,\frac{t^3}
        {\sqrt{t-t_0}} \, dt, = \taul\,{\cal F}_0\,\times 
	\frac{4}{7}\,\sqrt{1-\wzz} + \frac{6}{7}\,\wzz\,\fiso(3).
\end{equation}

Note that increasing values of $m$ also result in profile shapes
of lower amplitude.  In fact it is possible to solve for the
integrated light from a resonance scattering line in some special
cases.  Defining

\begin{equation}
F_{\rm tot}(m) = \int_{-1}^{+1}\,\fiso\,d\wz,
\end{equation}

\noindent a few selected results are $F_{\rm tot}(1)=4\pi\cdot\taul{\cal
F}_0$, $F_{\rm tot}(2) = 5\pi/3\cdot\taul\,{\cal F}_0$, 
$F_{\rm tot}(3)=2\pi/3\cdot\taul\,{\cal F}_0$, and
$F_{\rm tot}(4)=\pi/2\cdot\taul\,{\cal F}_0$.

It would appear that with $\taul=constant$, different values of
$F_{\rm tot}$, and thus lines of different brightness levels, could
result.  This seems counterintuitive if lines characterized by
different values of $m$ have the same optical depth.  However,
$\taul$ depends only on the density at the inner portion of the
disk $\Sigma_0$, not the value of $m$.  Thus when the line is
optically thin, the appropriate optical depth to use is one that
is angle averaged, similar in spirit to $\bar{\tau}$ in Brown \&
McLean (1977) for optically thin electron scattering.  Hence use
of a new optical depth parameter $T_l = \tau_0\times \tau_0 {\cal
F}_0/F_{\rm tot}(m)$ would ensure that lines of different $m$ values
with {\em isotropic} scattering will have the same total line
emission (i.e., ``area under the curve'') even though they have
different profile shapes.

\section{The Case of Resonance Scattering with $B=0$}	\label{app:Bzero}

For resonance line scattering with $E_1 \neq 0$ but with $B=0$
everywhere, the vector scattering function $\vec{h}$ greatly
simplifies.  Following Paper~II, we have that $\delta=\varphi$,
$\alpha_2 = 0$, $C=\cos 2\varphi$, $D=\sin 2\varphi$, and $\psis =
0$.  The components of the phase function become

\begin{equation}
\hanvec = \left\{ \begin{array}{lcl} 
	h_I & = & 1+\slantfrac{3}{8}E_1\,\left[\slantfrac{1}{3}\,\left(1-3\cos^2 i
		\right) + \sin^2 i\,\cos 2\varphi \right] \label{eq:hIzero} \\
	h_Q & = & \slantfrac{3}{8}E_1\,\left[\sin^2 i
		-\left(1+\cos^2 i \right)\,\cos 2\varphi \right] \label{eq:hQzero} \\
	h_U & = & \slantfrac{3}{8}E_1\,\cos i\,\sin 2\varphi. \label{eq:hUzero}  
	\end{array} \right.
\end{equation}

\noindent The Stokes flux of scattered light is

\begin{equation}
\vec{\cal F}^{\rm sc} = \taul\,{\cal F}_0\times 2\int_{t_0}^1\,\frac{t^{m-1}}
        {\sqrt{t-t_0}} \, \vec{h}\,dt.
\end{equation}

\noindent Note that care must be taken in dealing with terms that
are odd and even in $\varphi$ between angles of 0 and $\pi$.  For
example, $h_U$ is odd around the loop, ensuring that ${\cal F}_U=0$.
Accounting for the odd/even terms, and using the fact that $\sin^2
\varphi =t_0/t=\wzz/t$, solutions for the scattered flux in Stokes-I
and Stokes-Q are:

\begin{eqnarray}
\FI^{\rm sc} & = & \left(1+\slantfrac{1}{8}E_1-\slantfrac{3}{8}E_1\cos 2i\right)\,
	\fiso (m)-\slantfrac{3}{4}E_1\,\sin^2 i\,t_0\,\fiso(m-1), \\
\FQ^{\rm sc} & = & -\slantfrac{3}{4}E_1\,\cos^2 i\,\fiso(m) +
	\slantfrac{3}{4}E_1\,\left(1+\cos^2 i\right)\,t_0\,\fiso(m-1).
\end{eqnarray}

\noindent Note at the extrema of the line wings, $\fiso(m)=\fiso(m')$
and $t_0=1$, and $\qs = 3E_1/(4-E_1)$ always.

The resultant polarization across the line is entirely in Stokes-Q
when the observer's reference axes are aligned with the symmetry
axis of the star.  With $E_1=0.5$,
Figure~\ref{figApp1} shows a plot of ${\cal F}_I^{\rm
sc}$ for $m=3$ (left) and $m=4$ (right) at different viewing
inclination angles of $i=0^\circ, 30^\circ, 45^\circ, 60^\circ,$ 
and $90^\circ$ along the top panels.
The profiles are normalized with respect to the total emission
produced if the line had been isotropically scattering.  At bottom
is the relative fractional polarization $\qs={\cal F}_Q^{\rm sc}/{\cal F}_I^{\rm
sc}$ for $m=3$ and $m=4$ for the same viewing inclinations.  Note
that at the edges of the line, $\qs = 3/7$ for $E_1=0.5$, independent
of the viewing inclination, as expected under the point star
approximation.  

\begin{figure}[t]
\plotone{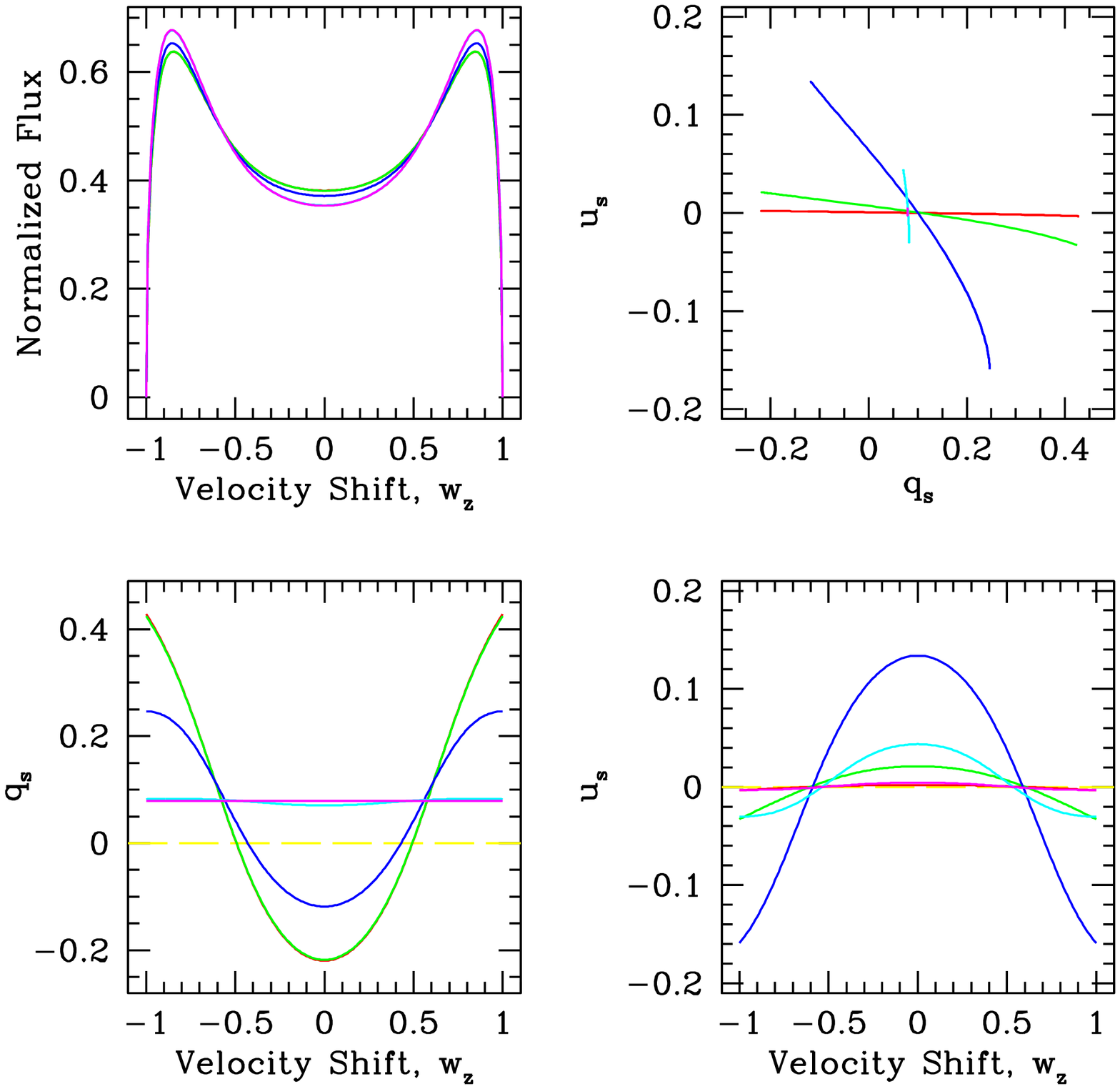}
\caption{Similar to Fig.~\ref{fig5}, here for the case of an axial
field in the point star approximation.  
\label{figApp2}}
\end{figure}

The goal here is to illustrate the polarimetric ``efficiency''.
The actual measured fractional polarization would be much smaller
owing to dilution by direct starlight.  These efficiencty curves
are relatively smooth functions of velocity shift.  This smoothness
is partly due to the fact that dipole scattering is a fairly slowly
varying function of location around the disk and also because
isovelocity loops sample a range of scattering angles.

\section{The Case of an Axial Field $B=B_{\rm Z_\ast}$}	\label{app:axial}

For an axial magnetic field with $\vec{B}=B_{\rm Z_\ast}(\varpi)\,
\hat{Z}_\ast$, we have that $\psis=0$ and $\thetas=i$.  The scattering
phase functions are quite similar to the zero field case, except
that now the Hanle effect appears in factors in the functions $C$
and $D$.  Using Paper~II, the phase functions are given by

\begin{equation}
\vec{h} = \left\{ \begin{array}{lcl} 
	h_I & = & 1+\slantfrac{3}{8}E_1\,\left[\slantfrac{1}{3}\,\left(1-3\cos^2 i
		\right) + \sin^2 i\,C(\varpi,\varphi) \right] \\
	h_Q & = & \slantfrac{3}{8}E_1\,\left[\sin^2 i
		-\left(1+\cos^2 i \right)\,C(\varpi,\varphi) \right] \\
	h_U & = & \slantfrac{3}{4}E_1\,\cos i\,D(\varpi,\varphi).  
	\end{array} \right. \label{eq:axial}
\end{equation}

\noindent where 

\begin{eqnarray}
C(\varpi,\varphi) & = & \phantom{-}\cos^2 \alpha_2\cos 2\varphi
	-\slantfrac{1}{2}\sin 2\alpha_2\sin 2\varphi \\
D(\varpi,\varphi) & = & -\cos^2\alpha_2\sin 2\varphi-\slantfrac{1}{2}\sin 2\alpha_2
	\cos 2\varphi,
\end{eqnarray}

\noindent where 

\begin{equation}
\cos^2 \alpha_2 = \frac{1}{1+b_0^2\,t^2},
\end{equation}

\noindent and

\begin{equation}
\cos \alt \,\sin \alt = \frac{b_0\,t}{1+b_0^2\,t^2}.
\end{equation}

Solutions for the vector Stokes flux is no longer analytic.  With
the Hanle effect, ${\cal F}_U \neq 0$ except for $B=0$ or in the saturated
limit.  In the latter case of $b_0 \gg 1$ everywhere, 
an analytic solution can be obtained,
which is given by

\begin{eqnarray}
{\cal F}_I^{\rm sc} & = & \left(1+\slantfrac{1}{8}E_1-\slantfrac{3}{8}E_1\cos 2i\right)\,
	\fiso (m)\\
{\cal F}_Q^{\rm sc} & = & \slantfrac{3}{4}E_1\,\sin^2 i\,\fiso(m).
\end{eqnarray}

\noindent Note that in this limit, the relative polarization becomes

\begin{equation}
\qs ={\cal F}_Q^{\rm sc}/{\cal F}_I^{\rm sc} = \frac{3E_1\sin^2 i}{8+E_1\,(1-3\cos^2 i)},
\end{equation}

\noindent which is a constant across the profile and
a function of viewing inclination only.  This means that the polarized
profile is flat-topped. 

Figure~\ref{figApp2} displays profiles for the axial field case
with $E_1=0.5$ at a fixed value of $\sin^2 i = 0.4$ but with different
field values at the inner disk radius of $\varpi=1$, of $\log b_0
= -2, -1, 0, +2, +4$ to achieve a large dynamic range in Hanle
ratios throughout the disk.  As $b_0$ increases, the location where
$B=\BHan$ moves outward to $\varpi = b_0$.  The upper left panel
in Figure~\ref{fig8} shows normalized profiles of ${\cal F}_I^{\rm
sc}$; lower panels show the fractional polarizations \qs\ and \us;
and upper right shows the \qs-\us\ shapes.  The color sequencing
is the same as in Figure~\ref{fig5}.  In large part the effect of
an axial field is to rotate and foreshorten the Q-U segments relative
to the zero field case.

\section{The Case of a Toroidal Field $B=B_\varphi$}	\label{app:toroidal}

For a toroidal magnetic field with $\vec{B}=B_\varphi\,\hat{\varphi}$,
the spherical geometry for the scattering problem is moderately
complex.  It is not possible to write down simple complete expressions
for the phase scattering functions.  But as demonstrated in Paper~II,
there are still some special cases that are analytic.  For example in the
saturated limit, ${\cal F}_U^{\rm sc} = 0$, and the $I$ and $Q$ fluxes
become

\begin{eqnarray}
{\cal F}_I^{\rm sc} & = & \left(1+\slantfrac{1}{8}E_1\right)\,\fiso(m)
	-\slantfrac{3}{8}E_1\sin^2 i\,t_0\,\fiso(m-1) \\
{\cal F}_Q^{\rm sc} & = & -\slantfrac{3}{8}E_1\,\left[\fiso(m) - (1+\cos^2 i)\,
	t_0\,\fiso(m-1)\right].
\end{eqnarray}

\noindent For a disk viewed edge-on, solutions for the Stokes fluxes
cannot be derived analytically; however, the scattering functions
simplify to

\begin{equation}
\vec{h} = \left\{ \begin{array}{lcl} 
	h_I & = & 1+\slantfrac{3}{8}E_1\,\left(1-3\sin^2 \varphi\right) 
		+ \slantfrac{3}{8}E_1\,\cos^2 \varphi\,\cos^2 \alpha_2, \\
	h_Q & = & -\slantfrac{3}{8}E_1\,\cos^2 \varphi + \slantfrac{3}{8}E_1\,
		(1+\sin^2\varphi)\,\cos^2 \alpha_2, \\
	h_U & = & -\slantfrac{3}{8}E_1\,\sin \varphi\,\sin 2\alpha_2.
	\end{array} \right.
\end{equation}

\begin{figure}[t]
\plotone{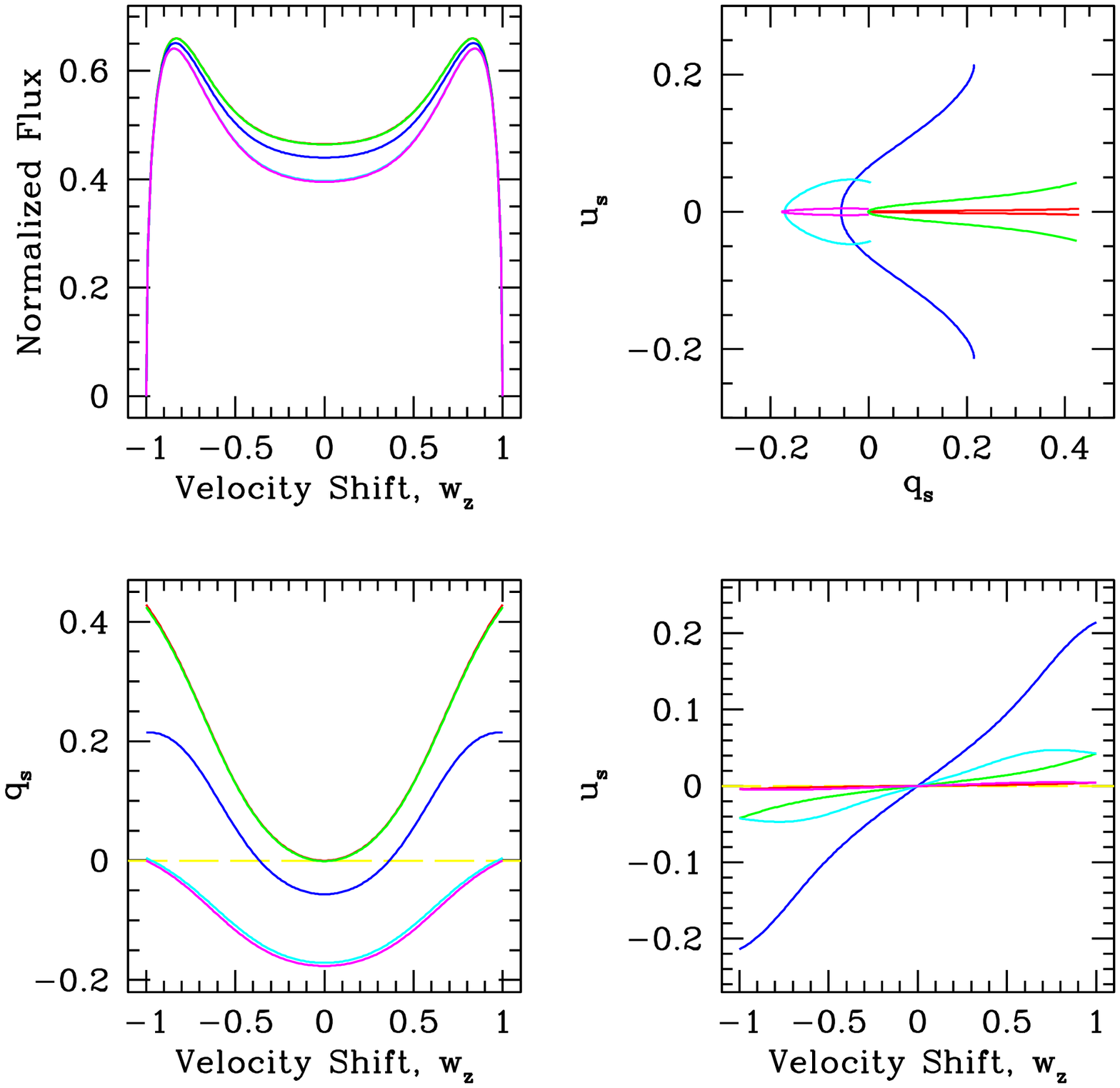}
\caption{Similar to Fig.~\ref{figApp2}, here for the case of a toroidal
field in the point star approximation.  
\label{figApp3}}
\end{figure}

\noindent Examples of scattering and polarized profiles are shown
in Figure~\ref{figApp3} as well as Q-U plots across the polarized
profiles.  In the previous section, $\sin^2 i=0.4$ was used for an
axial field to give a significant signal in \us.  For this toroidal
field case, an edge-on disk with $\sin^2 i=1.0$ was used.  The style
of this figure is the same as Figure~\ref{figApp2}.  The a major
distinctive in relation to a disk with an axial field is that a
toroidal field leads to \us\ profiles that are antisymmetric instead
of symmetric.

\end{document}